\documentclass[aps,prl,floatfix,twocolumn,superscriptaddress,showpacs,reprint]{revtex4}

\usepackage{graphicx}
\usepackage{mathrsfs}
\usepackage{amsmath}
\usepackage{amsfonts}
\usepackage{subfigure}
\usepackage{bm}
\usepackage{verbatim}
\usepackage{color}
\usepackage{xcolor}
\usepackage[colorlinks=true, letterpaper=true, pdfstartview=FitV, linkcolor=blue, citecolor=blue, urlcolor=blue]{hyperref}
\usepackage{siunitx}

\hypersetup{%
pdftitle={Many body effects for quasirelativistic Dirac Fermions at finite momentum density in the clean limit},%
pdfauthor={Inti Sodemann},%
pdfpagemode={UseNone},%
pdfstartview={FitH},%
breaklinks=true}

\newcommand{\be}{\begin{equation}}
\newcommand{\ee}{\end{equation}}
\newcommand{\ba}{\begin{equation}\begin{split}}
\newcommand{\ea}{\end{split}\end{equation}}

\begin{document}
\title{Current-induced and interaction-driven Dirac point drag of massless quasirelativistic Fermions}
\author{Inti Sodemann}
\affiliation{Department of Physics, Massachusetts Institute of Technology, Cambridge, Massachusetts 02139}

\begin{abstract}
We study the quasiparticle properties of two-dimensional massless Dirac Fermions when the many-body states possess a finite momentum density in the clean limit. The lack of Galilean invariance endows the many-body states at finite momentum density with qualitative differences from those of the system at rest. At finite carrier densities we demonstrate the appearance of a current-induced distortion of the pseudospin texture in momentum space that can be viewed as a drag of the Dirac point and the origin of which lies entirely in electron-electron interactions. We discuss the potential observation of this effect in graphene.  
\end{abstract}
\maketitle

\noindent
{\color{blue}{\em Introduction}}---Momentum is not often regarded as a good conserved quantity in electronic systems due to the presence of disorder and momentum transfer from electronic degrees of freedom to the lattice. However, in recent years, the advancement in quality in selected materials has made increasingly relevant the case for including the momentum conservation in our picture of electron transport. This regime, often referred to as hydrodynamic, has received considerable theoretical attention in graphene~\cite{Muller08,Fritz08,Muller09,Foster09,Narozhny15,Levitov15,Principi15,Lucas15}, and recent experiments have found evidence for hydrodynamic transport in high-quality samples~\cite{Geim,Crossno2014}. 

In this work we will focus on the zero-temperature limit of Dirac Fermions with a finite carrier density. The momentum conservation allows one to consider the ground states in different subspaces of the many-body Hilbert space that differ by their total momentum density. In a Galilean invariant system these different subspaces can be mapped into one another by a Galilean boost, and therefore there is no new physics to the problem at finite momentum density. In a quasirelativistic system like graphene, the kinetic energy is Lorentz invariant but the interactions are not, and therefore there exists no simple mapping between the problem at different momentum densities making it a nontrivial parameter in the problem.

We will develop a simple mean field theory of the quasiparticle properties of a system of interacting massless Dirac Fermions system, such as those arising in graphene, and show that in order to minimize their exchange energy in the current-carrying many-body state the electrons pseudospin orientation changes in a way that can be described as a current-induced {\it drag} of the Dirac the point.

\noindent
{\color{blue}{\em noninteracting limit with finite momentum density}}---Consider a system of noninteracting two-dimensional massless Dirac Fermions. We would like to find its ground state under the constraints of a given total electron number and momentum:

\be
N=\int d^2r \ \psi^{\dagger}_r \psi_r, \ {\mathbf P}=\int d^2r \ \psi^{\dagger}_r {\mathbf p}  \psi_r.
\ee

\noindent In order to obtain states that have a finite momentum and particle density in the thermodynamic limit we introduce Lagrange multipliers $\mu$ and ${\mathbf u}$ and find the unconstrained ground state of the following free energy: 

\begin{equation}\label{xx}
\begin{split}
F&=H-\mu N-{\mathbf u}\cdot {\mathbf P}=\int d^2r \ \psi^{\dagger}_r h \ \psi_r , \\
h&=v \sigma \cdot {\mathbf p}-\mu-{\mathbf u}\cdot {\mathbf p}.
\end{split}
\end{equation}


\noindent where $\sigma$ denotes the Pauli matrices in the pseudospin space~\footnote{In graphene the pseudospin labels sublattice degrees of freedom while in topological insulators it labels real spin.}. At finite carrier density the Fermi surface can be shown to be an ellipse described in polar coordinates by:

\be
p_F(\theta)=\frac{|\mu|}{v-s u \cos\theta}.
\ee

\noindent Here $\theta$ is the polar angle measured from the axis defined by $\mathbf{u}$, and $s=1(-1)$ for electrons (holes). The particle number ($n$) and current densities (${\mathbf j}$) can be found to be:

\begin{equation}\label{n&j}
\begin{split}
&n \equiv \int _{ h < 0}\frac{d^2p}{(2 \pi )^2}  =n_D+\frac{s \mu^2}{4 \pi v^2}\frac{1}{(1-\beta^2)^{3/2}}, \\
&{\mathbf j} \equiv \int _{ h < 0}\frac{d^2p}{(2 \pi )^2} v  \langle \sigma \rangle=\mathbf{u}\frac{s \mu^2}{4 \pi v^2}\frac{1}{(1-\beta^2)^{3/2}},
\end{split}
\end{equation}

\noindent where $\beta=u/v$, $n_D$ is the density at the Dirac point. 
Notice that ${\mathbf j} =(n-n_D)\mathbf{u}$, making manifest the interpretation of $\mathbf{u}$ as the average velocity for charge transport. The current density will vanish at the Dirac point. As we will see, this conclusion follows from electron-hole symmetry even in the presence of interactions. This is an striking property of Dirac Fermions at the Dirac point, that, even when they are unable to relax their total momentum, they can reach an equilibrium state with zero total current. This is impossible for Galilean Fermions where a finite momentum density is always accompanied by a finite current density.  

{\color{blue}{\em Symmetry considerations}}---We briefly describe in this section the constraints imposed by certain symmetries of the problem. Consider the following operations:

\begin{equation}\label{xx}
\begin{split}
T \psi_r T^{-1}&=i \sigma_y \psi_r, \ T i T^{-1}=-i, \\
C \psi_r C^{-1}&=\sigma_x \psi_r^\dagger, \\
S \psi_r S^{-1}&=\sigma_z \psi_{-r}. 
\end{split}
\end{equation}

All of the above are expected to be good symmetries of the full Hamiltonian including interactions. They can be interpreted as a particle-hole conjugation ($C$), a time-reversal-like ($T$) and space-inversion-like ($S$) operations~\footnote{In graphene the conventional definition of time reversal and space inversion also exchange valleys, so, the symmetries here considered can be viewed as a composition with a suitable valley swap operation.}. The particle number and total momentum are however not invariant under these symmetries and their transformation properties are summarized in Table~\ref{tab}. The momentum is invariant under $C$, and the density is invariant under $C$ only at the Dirac point. However the current is odd under $C$. This implies that, at the Dirac point, the current must vanish even when averaged over a subspace of the Hilbert space with a definite momentum. This shows why current-carrying many-body state are not allowed at equilibrium even if the system has a net momentum. This fact originates physically from the property that the group velocity of a single particle state in the conduction band at momentum $\mathbf{p}$ points in the opposite direction to that of a state in the valence band with the same momentum. Therefore, particle-hole excitations connecting those states can change the current without changing the momentum. Away from the Dirac point the only symmetry that leaves the particle and momentum densities unchanged is the product $TS$, but the current is even under this transformation, hence quasiequilibrium current-carrying states at non-zero momentum are allowed.
 
{\color{blue}{\em Mean-field theory of interacting current-carrying states}}---In addition to the kinetic energy we consider an interaction term in the Hamiltonian:

\be
V=\frac{1}{2A}\sum_{qp p' \alpha\beta} v_q \psi_{p+q,\alpha}^\dagger\psi_{p'-q,\beta}^\dagger\psi_{p',\beta}\psi_{p,\alpha},
\ee

\noindent where $\alpha,\beta$ are labels for the pseudospin degree of freedom of the Dirac Fermion. For the Coulomb interaction we would have $v_q=2 \pi e^2/\epsilon q$, and $v_{q=0}=0$ from the neutralizing background. Let us consider the case of a finite density of holes~\footnote{The electron case is equivalent but we find slightly simpler to set up the notation for the hole side. The final Eqs.~\eqref{K}-\eqref{epsilon} can be used for electron ($\mu>0$) and hole ($\mu<0$) dopings.} and assume that they form a many-body Slater determinant in which every momentum eigenstate is either empty or singly occupied:

\be
|\Psi\rangle=\prod_{p \in FS} \biggl(\sum_\alpha u_{p\alpha} \psi^\dagger_{p \alpha}\biggr) |O\rangle,
\ee

\noindent Where $FS$ is the region in momentum that is singly occupied, and $u_{p\alpha}$ are the spinor coordinates parametrizing the orientation of the state occupied at momentum $p$ in the pseudospin Bloch sphere. We will minimize the energy under the constraint of fixed particle number and momentum within this set of states, therefore our procedure can be viewed as a form of Hartree-Fock theory at finite momentum. The free energy including interactions is:

\begin{equation}\label{xx}
\begin{split}
&F=\sum_{p}   {\rm tr}[G_p (v  \sigma\cdot {\mathbf p}-\mu-{\mathbf u}\cdot {\mathbf p}+\Sigma_p/2)], \\
&G_p\equiv f_p | {\mathbf n}_p\rangle \langle {\mathbf n}_p |, \ \Sigma_p\equiv -\frac{1}{A}\sum_{p'} v_{p-p'}G_{p'}.
\end{split}
\end{equation}

\noindent where $f_p=0(1)$ if the state is empty (occupied) and $| {\mathbf n}_p\rangle$ is the state corresponding to unit vector ${\mathbf n}_p$ in the pseudospin Bloch sphere that is occupied at momentum $\mathbf{p}$. More explicitly, the free energy reads as:

\begin{equation}\label{xx}
\begin{split}
&F=\sum_{p} (v  {\mathbf p}\cdot {\mathbf n}_p-\mu-{\mathbf u}\cdot {\mathbf p})f_p \\
&\cdots-\frac{1}{2 A}\sum_{p,p'}v_{p-p'} \left(\frac{1+{\mathbf n}_p \cdot {\mathbf n}_p'}{2}\right)  f_p f_p'.
\end{split}
\end{equation}

\begin{table}
\caption{Action of time-reversal ($T$), charge-conjugation ($C$) and space-inversion ($S$) on the density measured from the Dirac point $n-n_D$, the current density ${\mathbf j}$, and the total many-body momentum ${\mathbf P}$.} 
\centering 
\begin{tabular}{c | c  c  c  c  c  c} 
\hline
\hline 
 & $T$ & $C$ & $S$ \\ [0.5ex] 
\hline 
$n-n_D$ & \ + & - & +  \\ [0.5ex]
${\mathbf j}$ & \ - & - & -  \\ [0.5ex] 
${\mathbf P}$ & \ - & + & -  \\ [0.5ex]
\hline\hline 
\end{tabular}
\label{tab} 
\end{table}


To gain insight into the problem it is useful to view it as a classical two-dimensional magnet. The momentum ${\mathbf p}$ would play the role of the real space site at which Heisenberg-like (${\mathbf n}_p$) and Ising-like ($f_p$) degrees of freedom reside. For purely repulsive interactions the Heisenberg pseudospins are coupled ferromagnetically, but there is a Zeeman-like field $v {\mathbf p}$ that tries to pin the pseudospins antiparallel to ${\mathbf p}$ and creates a vortex-like configuration the singularity of which is the Dirac point. Variations of the energy functional with respect to $f_p$ can be written as:

\begin{equation}\label{E}
\begin{split}
&\delta_{f_p} F=\sum_{p} \varepsilon_p \delta f_p, \\
&\varepsilon_p \equiv v  {\mathbf n}_p \cdot {\mathbf p}-\mu-{\mathbf u}\cdot {\mathbf p}-\frac{1}{A}\sum_{p'}v_{p-p'} \left(\frac{1+{\mathbf n}_p \cdot {\mathbf n}_p'}{2}\right) f_p'. 
\end{split}
\end{equation}

\noindent Demanding $\delta_{f_p} F$ to be non-negative determines the shape of the Fermi surface: $f_p=1-\theta(\varepsilon_p)$. To vary $F$ with respect to ${\mathbf n}_p$ we add a set of Lagrange multipliers to enforce its unit length constraint: $F+\sum_p \lambda_p /2 {\mathbf n}_p^2$. The variation is:

\begin{equation}\label{K}
\begin{split}
&\delta_{{\mathbf n}_p} F=\sum_{p} [v ({\mathbf p}-{\mathbf K}_p)+\lambda_p {\mathbf n}_p] \cdot \delta {\mathbf n}_p, \\
&v {\mathbf K}_p \equiv \frac{1}{2 A}\sum_{p'}v_{p-p'}  {\mathbf n}_{p'} f_{p'}. 
\end{split}
\end{equation}

\noindent Demanding this variation to be zero provides the following equation for the unit vector:

\be\label{n}
{\mathbf n}_p=-\frac{{\mathbf p}-{\mathbf K}_p}{|{\mathbf p}-{\mathbf K}_p|}, 
\ee

\noindent Equations~\eqref{E}-\eqref{n} define a self-consistent loop determining the dispersion and the coherent combination of states being occupied. Substituting Eq.~\eqref{n} into Eq.~\eqref{E} leads to a more succinct expression for the dispersion:  

\be
\varepsilon_p = -v |{\mathbf p}-{\mathbf K}_p|-\mu-{\mathbf u}\cdot {\mathbf p}-\frac{1}{2 A}\sum_{p'}v_{p-p'} f_{p'}.
\ee

\noindent The vector ${\mathbf K}_p$ is of central importance to this work. It is the same term responsible for the logarithmic corrections to the quasiparticle dispersion at the Dirac point when the system carries no current~\cite{Gonzalez94}. More crucially, ${\mathbf K}_p$ determines the location in momentum space of the Dirac point: ${\mathbf p}_D={\mathbf K}_{p_D}$. In the absence of current the self-consistent solution to this equation will be ${\mathbf p}_D=0$ but in the current-carrying many-body state the fact that ${\mathbf K}_{{\mathbf p}=0}\neq 0$ shows that the Dirac point is displaced in momentum, therefore the current induces a {\it drag} in the Dirac point. 

\begin{figure}
\begin{center}
\includegraphics[width=3in]{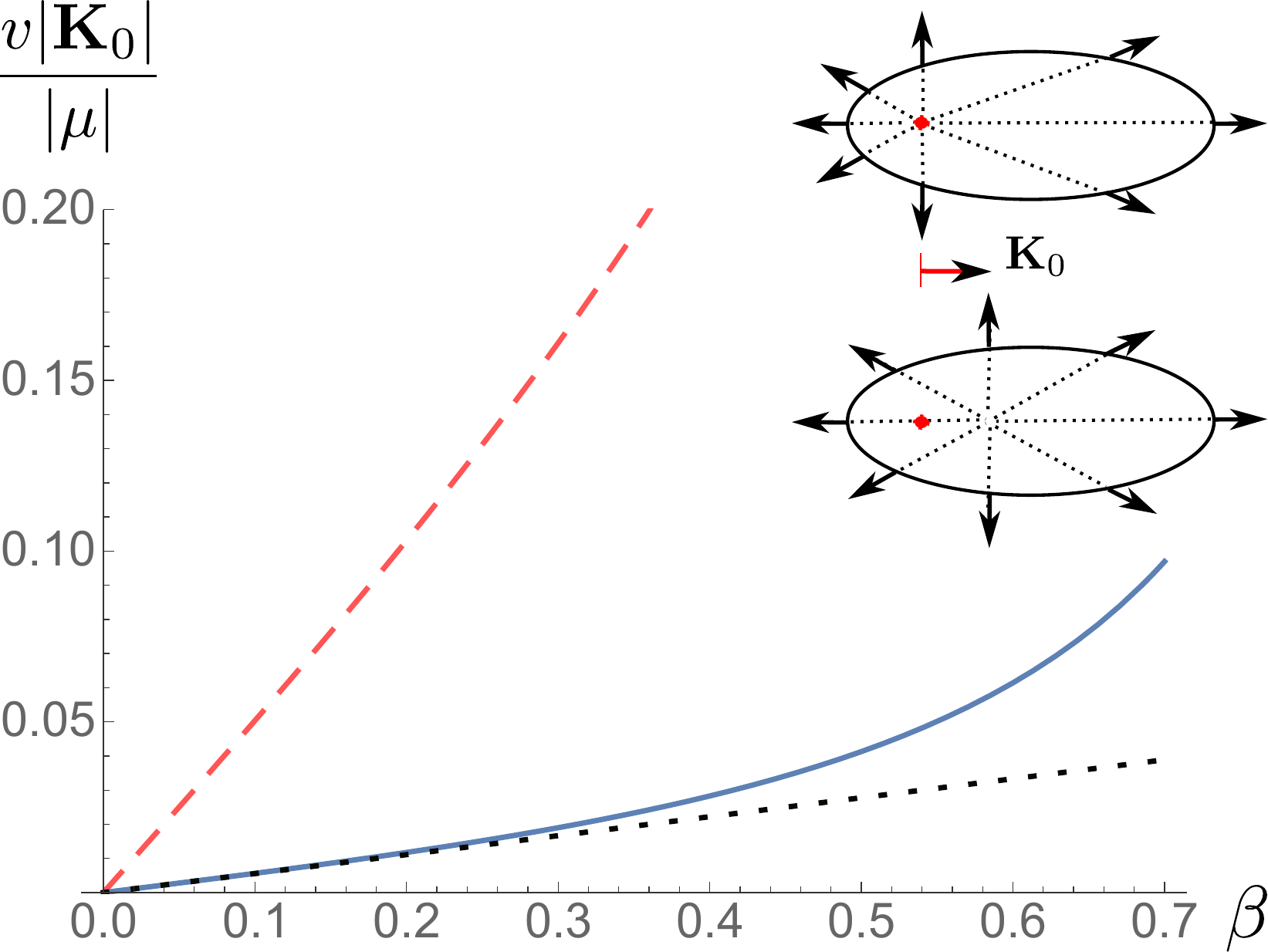}
\caption{(color online) Dirac-point drag magnitude as a function of the average charge transport velocity $\beta = u/v$. The solid line is the prediction for the model screened interaction in graphene and the dashed line is for the bare Coulomb interaction. The dotted line is the linear in $\beta$ approximation from Eq.~\eqref{Kapprox}. Upper inset: pseudospin orientation near the elliptical Fermi surface in the noninteracting current-carrying state. Lower inset: rearrangement of the pseudospins produced by interactions that accompanies the Dirac point drag. The red dot indicates the origin in momentum, ${\mathbf p}=0$, and the Dirac point is located where the dotted lines meet.}
\label{Kvsbeta}
\end{center}
\end{figure}

The Dirac point drag originates as a way for the system to save on the exchange energy cost imposed by the pseudospin vortex in momentum space. At zero current the Fermi surface symmetrically surrounds the Dirac point and the pseudospin vortex core is pinned at its center at zero momentum. In the noninteracting limit, when the system is in a current-carrying many-body state, the Fermi surface is deformed and its boundary in momentum space becomes closer to the vortex core. In the hole-doping case the states inside the Fermi surface are unoccupied, and the occupied states closer to the Fermi surface are responsible for most of the exchange energy cost. Therefore, when interactions are turned on the vortex core is pushed further into the center of the region of un-occupied states to save some of the exchange energy cost by making the pseudospins of states near the Fermi surface closer to parallel. This picture is illustratred in the inset to Fig.~\ref{Kvsbeta}. 

In the case of graphene we have a Dirac Fermion degeneracy of four, accounting for valley and spin multiplicities, and the picture we just described would be essentially replicated for each of these Dirac Fermions. Also, we would like to note in passing that the coupling of the electron pseudospin to the current-density fluctuations in graphene has been studied near equilibrium and shown to lead to an interesting enhancement of the Drude peak and the plasma frequencies which are also related to its lack of Galilean invariance~\cite{Polini}. 

{\color{blue}{\em Perturbative estimates for the Dirac point drag}}---In this section we will compute perturbatively in the strength of the Coulomb interaction the Dirac point drag. In order to assess the impact of screening we employ the following simplified model for the screened Coulomb interaction:

\be
v_q=\frac{2 \pi e^2}{q+2 \pi e^2 \nu_F}
\ee

\noindent where $\nu_F=g p_F/(2 \pi v)$ is the density of states of the noninteracting system at zero current, and $g=4$ accounts for the spin-valley degeneracy. To estimate ${\mathbf K}_p$ peturbatively we evaluate it from Eq.~\eqref{K} replacing in the right hand side the Fermi surface and the pseudospin orientation of the noninteracting system: ${\mathbf n}_p^0=-\hat{\mathbf{p}}$. We obtain thus the Dirac point drag to first order in the screened Coulomb interaction, ${\mathbf K}_0\equiv{\mathbf K}_{p=0}$, to be:

\begin{equation}\label{K0}
\begin{split}
&{\mathbf K}_{0} = \frac{\alpha \mu \hat{{\mathbf u}}}{2 u} \times \\
&\left[\frac{1}{\sqrt{1-\beta^2}}+\alpha  g \sqrt{1-\beta^2}-\sqrt{(1+\alpha  g)^2-(\alpha  g \beta )^2}\right],
\end{split}
\end{equation}

\noindent where $\alpha=e^2/v$ is the effective fine structure constant of graphene. The perturbative expression in the case for the bare Coulomb interaction can be conveniently obtained from that above by taking $g\rightarrow 0$. Figure~\ref{Kvsbeta} depicts the behavior of the Dirac point drag as a function of $\beta$. The following is a good linear in $\beta$ approximation to Eq.~\eqref{K0}:

\be\label{Kapprox}
{\mathbf K}_{0}\approx   \frac{\alpha \beta \mu \hat{{\mathbf u}}}{4v (1+g \alpha)}+\mathcal{O}(\beta^2).
\ee

Another quantity of interest is the energy at the Dirac point in the current-carrying many-body state which can be estimated to first order in the screened Coulomb interaction to be: 

\begin{equation}\label{epsilon}
\begin{split}
&\varepsilon_0 \equiv\varepsilon_{{\mathbf p}_D}+\mu+{\mathbf u}\cdot {\mathbf p}_D=-\frac{1}{2 A}\sum_{p'}v_{{\mathbf p}_D-p'} f_{p'}, \\
&\varepsilon_0 \approx -\frac{U_0}{2}-\frac{\alpha \mu}{2}\times \cdots \\
&\left[\frac{1}{\sqrt{1-\beta^2}}+\alpha g \log \left(\frac{ g\alpha +g\alpha \sqrt{1-\beta ^2}}{1+g \alpha+\sqrt{(1+ g \alpha)^2-(g \alpha  \beta)^2}}\right)\right],
\end{split}
\end{equation}

\noindent where $U_0=1/A\sum_{p}v_{p}$ is a Hubbard-type on-site energy scale which is independent of $\mu,u$. To leading order in the interaction strength we can re-express $\varepsilon_0$ and ${\mathbf K}_{0}$  in terms of the density $n$ and the current density ${\mathbf j}$, which are quantities directly accessible to experiment, by using the noninteracting expressions in Eq.~\eqref{n&j}.

At a fixed current density the parameter $\beta$ increases as the density approaches the Dirac point as $\sim \frac{1}{|n-n_D|}$. However, in realistic graphene samples disorder-induced charge inhomogeneities become more prominent as the Dirac point is approached~\cite{Yacoby,Crommie}. Assuming $|n-n_D|\gtrsim 10^{12}$ cm$^{-2}$, which should be sufficient to ignore charge fluctuations in high-quality samples such as those on boron-nitride substrates~\cite{Pablo}, and a current of $I=1$mA traversing a 1-$\mu$m -wide sample, leads to an estimate $\beta \lesssim 0.6$. From Eq.~\eqref{K} this leads to an estimate of $|{\mathbf K}_{0}|\sim 10^{-3}$ \AA$^{-1}$. Therefore the effect is small but perhaps within reach of high resolution angle-resolved-photoemission-spectroscopy if it could be realized for samples in the presence of large current densities. 

Finally, we wish to emphasize that the picture we described requires essentially a {\it local} validity of the hydrodynamic description, not a global one. In a sample with ``clean" regions the electron-electron collision mean free path could be smaller than the impurity or phonon collisions mean free paths, hence validating hydrodynamics locally. The feasibility of achieving this regime has been recently demonstrated in experiments~\cite{Geim,Crossno2014}. Thus, in the steady state of current flow there would be ``clean" regions in the hydrodynamic regime with local values of the thermodynamic potentials $\mu$ and ${\bf u}$. Provided that the variation of such quantities is sufficiently smooth on the scale of the Fermi wavelength we expect our picture to hold. Local spectroscopic measurements in such clean regions could examine the drag of the Dirac point we describe.    

{\color{blue}{\em Summary}}---We have described a many body approach to the current-carrying many-body states in the clean limit that is nonperturbative in the current but relies on the conservation of momentum for applying a quasiequilibrium treatment. More specifically, we have studied quasiparticle self-energy effects in the current-carrying many-body states of interacting massless Dirac Fermions. An interesting drag of the Dirac point arises as a mean for the system to reduce exchange energy. Although numerically small, the effect might be observable in high quality graphene samples in the regime of large current densities and small carrier densities and sufficiently away from the Dirac point to ignore disorder-induced charge inhomogeneities.

 

{\color{blue}{\em Acknowledgements}}---
I would like to thank B. Skinner and L. Levitov for stimulating discussions. IS is supported by the Pappalardo Fellowship.

\bibliographystyle{apsrev4-1}

\begin{thebibliography}{100}

\bibitem{Muller08} M. Muller, L. Fritz, and S. Sachdev, Phys. Rev. B {\bf 78}, 115406 (2008).

\bibitem{Fritz08} L. Fritz, J. Schmalian, M. Muller, and S. Sachdev, Phys. Rev. B {\bf 78}, 085416 (2008).

\bibitem{Muller09} M. Muller, J. Schmalian, and L. Fritz, Phys. Rev. Lett. {\bf 103}, 025301 (2009).

\bibitem{Foster09} M. S. Foster and I. L. Aleiner, Phys. Rev. B {\bf 79}, 085415 (2009).

\bibitem{Narozhny15} B. N. Narozhny, I. V. Gornyi, M. Titov, M. Schutt, and A. D. Mirlin, Phys. Rev. B {\bf 91}, 035414 (2015).

\bibitem{Levitov15} L. Levitov, G. Falkovich, arXiv:1508.00836 (2015).

\bibitem{Principi15} A. Principi, G. Vignale, M. Carrega, M. Polini, Phys. Rev. B {\bf 93}, 125410 (2016).

\bibitem{Lucas15} A. Lucas, J. Crossno, K. C. Fong, P. Kim, S. Sachdev, Phys. Rev. B {\bf 93}, 075426 (2016).

\bibitem{Geim} D. A. Bandurin, I. Torre, R. Krishna Kumar, M. Ben Shalom, A. Tomadin, A. Principi, G. H. Auton, E. Khestanova, K. S. Novoselov, I. V. Grigorieva, L. A. Ponomarenko, A. K. Geim, M. Polini, Science {\bf 351}, 1055 (2016).

\bibitem{Crossno2014} J. Crossno, J. K. Shi, K. Wang, X. Liu, A. Harzheim, A. Lucas, S. Sachdev, P. Kim, T. Taniguchi, K. Watanabe, T. A. Ohki, K. C. Fong, Science {\bf 351}, 1058 (2016).

\bibitem{Gonzalez94} J. Gonz\'alez, F. Guinea, M. A. H. Vozmediano, Nucl. Phys. B {\bf 424}, 595 (1994).

\bibitem{Polini}  S. H. Abedinpour, G. Vignale, A. Principi, M. Polini, W. K. Tse, and A. H. MacDonald, Phys. Rev. B {\bf 84}, 045429 (2011).

\bibitem{Yacoby} J. Martin, N. Akerman, G. Ulbricht, T. Lohmann, J. H. Smet, K. von Klitzing and A. Yacoby, Nature Phys. {\bf 4}, 144 (2008).

\bibitem{Crommie} Y. Zhang, V. W. Brar, C. Girit, A. Zettl and M. F. Crommie, Nature Phys. {\bf 5}, 722 (2009). 

\bibitem{Pablo} J. Xue, J. Sanchez-Yamagishi, D. Bulmash, P. Jacquod, A. Deshpande, K. Watanabe, T. Taniguchi, P. Jarillo-Herrero and B. J. LeRoy, Nature Mater. {\bf 10}, 282 (2011).

\end{thebibliography}

\end{document}